\documentclass[prl,twocolumn,aps,superscriptaddress]{revtex4-1}
\usepackage{color,epsfig}
\usepackage{bm}
\usepackage{amsmath,amsfonts,amssymb,bm}
\usepackage{hyperref}
\usepackage{subfigure}

\newcommand{\bea}{\begin{eqnarray}}
\newcommand{\eea}{\end{eqnarray}}
\newcommand{\be}{\begin{equation}}
\newcommand{\ee}{\end{equation}}

\renewcommand\vec{\bm}

\begin{document}

\title{Dislocation defect as a bulk probe of monopole charge of multi-Weyl semimetals}
\author{Rodrigo Soto-Garrido}
\affiliation{Facultad de F\'isica, Pontificia Universidad Cat\'olica de Chile, Vicu\~{n}a Mackenna 4860, Santiago, Chile}
\author{Enrique Mu\~{n}oz}
\affiliation{Facultad de F\'isica, Pontificia Universidad Cat\'olica de Chile, Vicu\~{n}a Mackenna 4860, Santiago, Chile}
\author{Vladimir Juri\v ci\' c}
\affiliation{Nordita, KTH Royal Institute of Technology and Stockholm University, Roslagstullsbacken 23, 10691 Stockholm, Sweden}

\begin{abstract}
Multi-Weyl semimetals  feature band crossings with the dispersion that is, in general, linear in only one direction, and as a consequence their band structure is characterized by the monopole charge $n$ which can be greater than one. We show that a single screw dislocation defect oriented in the direction connecting the nodal points, which acts as an effective pseudo-magnetic flux tube, can serve as a direct probe of the monopole charge $n\geq1$ characterizing the bulk band structure of a multi-Weyl semimetal. To this end, as a proof of principle, we propose a rather simple mesoscopic setup in which the monopole charge leaves a direct imprint on the conductance measured in the plane perpendicular to the dislocation. In particular, the ratio of the positions of the neighboring maxima in the conductance as a function of the gate voltage can serve to deduce the monopole charge, while the value of the effective pseudo-magnetic flux can be extracted from the position of a  conductance maximum. We expect that these findings will prompt further studies on the role of multiple dislocations, as well as other topological lattice defects, such as grain boundaries and disclinations, in topological nodal materials. 
\end{abstract}

\maketitle

\begin{figure}[t]
\centering
\includegraphics[width=80mm]{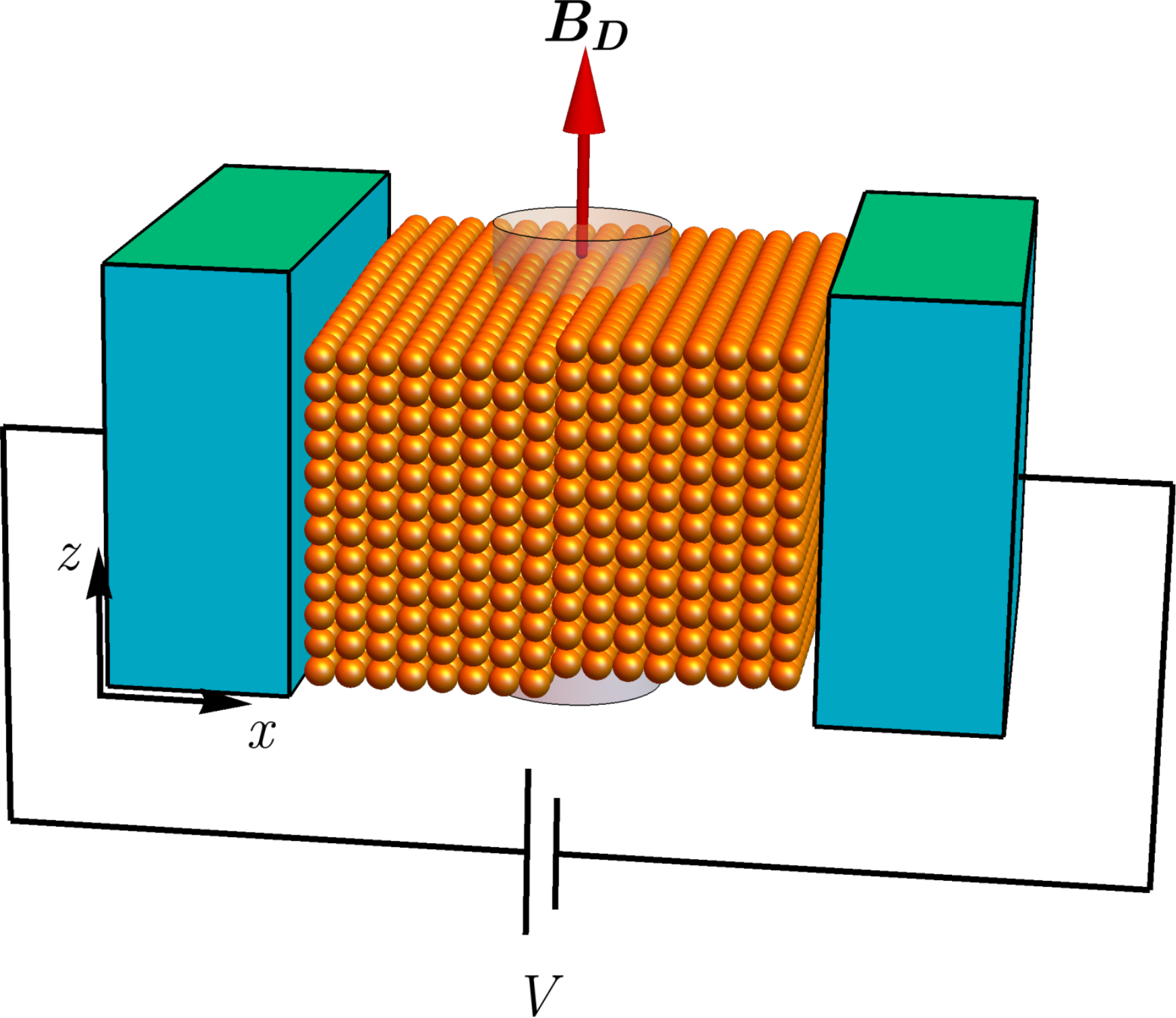}
\caption{Illustration of the setup for the transport experiment probing the monopole charge in a multi-Weyl semimetal.
The screw dislocation is characterized by a Burgers vector along the $z$ axis generating
an effective pseudo-magnetic field $B_D=\Phi_D/\pi a^2$ pointing in the same  direction as the Burgers vector.
The radius of the dislocation core is $a$. The leads are placed in the $x-z$ plane and are biased at a voltage $V$. The monopole charge and the effective flux are directly imprinted in the measured dependence of the conductance on the gate voltage, see Fig.~\ref{fig:conductances}. }
\label{fig:setup}
\end{figure}

{\it Introduction.~} Weyl semimetals are in focus of research in the condensed matter physics due to their exotic properties, such as unusual Fermi arc surface states and the chiral anomaly, intimately related to their topological nature~\cite{Hasan_017,armitage2017}. 
Since at least time-reversal or inversion symmetry is broken in these systems, valence and conduction bands can cross at  pairs of nodal points, representing the sources and the sinks of the Abelian Berry curvature in the Brillouin zone (BZ), which are characterized by the monopole charge $n$. In  conventional (single) Weyl semimetals $n=1$ while in multi-Weyl semimetals (mWSMs) can be greater than one with the crystalline symmetries bounding its maximum value to three \cite{XuPRL2011,FangPRL2012,BohmYang2014}.
As a consequence, close to these nodal points mWSMs host low-energy quasiparticles with the dispersion which is, in general, linear only in one direction, implying anomalous features in the transport \cite{ParkS2017,Sukhachov2017,Dantas2018,Lepori2018,nag2018,Sengupta2019,Dantas2019}.
Furthermore, by virtue of the bulk-boundary correspondence, these nodal points are connected via the $n$ topologically protected localized Fermi arc surface states. On the other hand, in crystalline solids dislocation defects are rather ubiquitous as they are energetically inexpensive. In this paper we propose that lattice dislocations can be used to directly probe the monopole charge in mWSMs in a simple mesoscopic setup shown in Fig.~\ref{fig:setup}.

The role of topological lattice defects has been extensively explored in gapped topological materials, such as topological insulators and superconductors, where, for instance, dislocations can host special topologically and symmetry protected modes~\cite{Vishwanath2009,teo2010,juricic2012,Asahi2012,Slager2013,Hughes2014,Slager2014,Chung2016,Teo2017,Slager2019,Tuegel2019}, the experimental signatures of which have been reported in Refs.~\cite{Hamasaki2017,Nayak2019}. 
Furthermore, these lattice defects in topological semimetals may serve as a platform for the realization of the chiral anomaly through protected gapless propagating  modes~\cite{Bulmash2015,Zubkov2015, Hiroaki2016,chernodub2017,Zemin2017,Zemin2019,Kodama2019}. Related insulating states, e.g. axion insulator, can also feature such defect modes~\cite{WangZhong2013,YizhiYou2016}. A natural question in this context is therefore what is the relationship between the topological features of a mWSM and lattice dislocations, in particular, whether these defects can probe the monopole charge in a mWSM.

\begin{figure*}[t!]
\centering
{\includegraphics[width=0.324\textwidth]{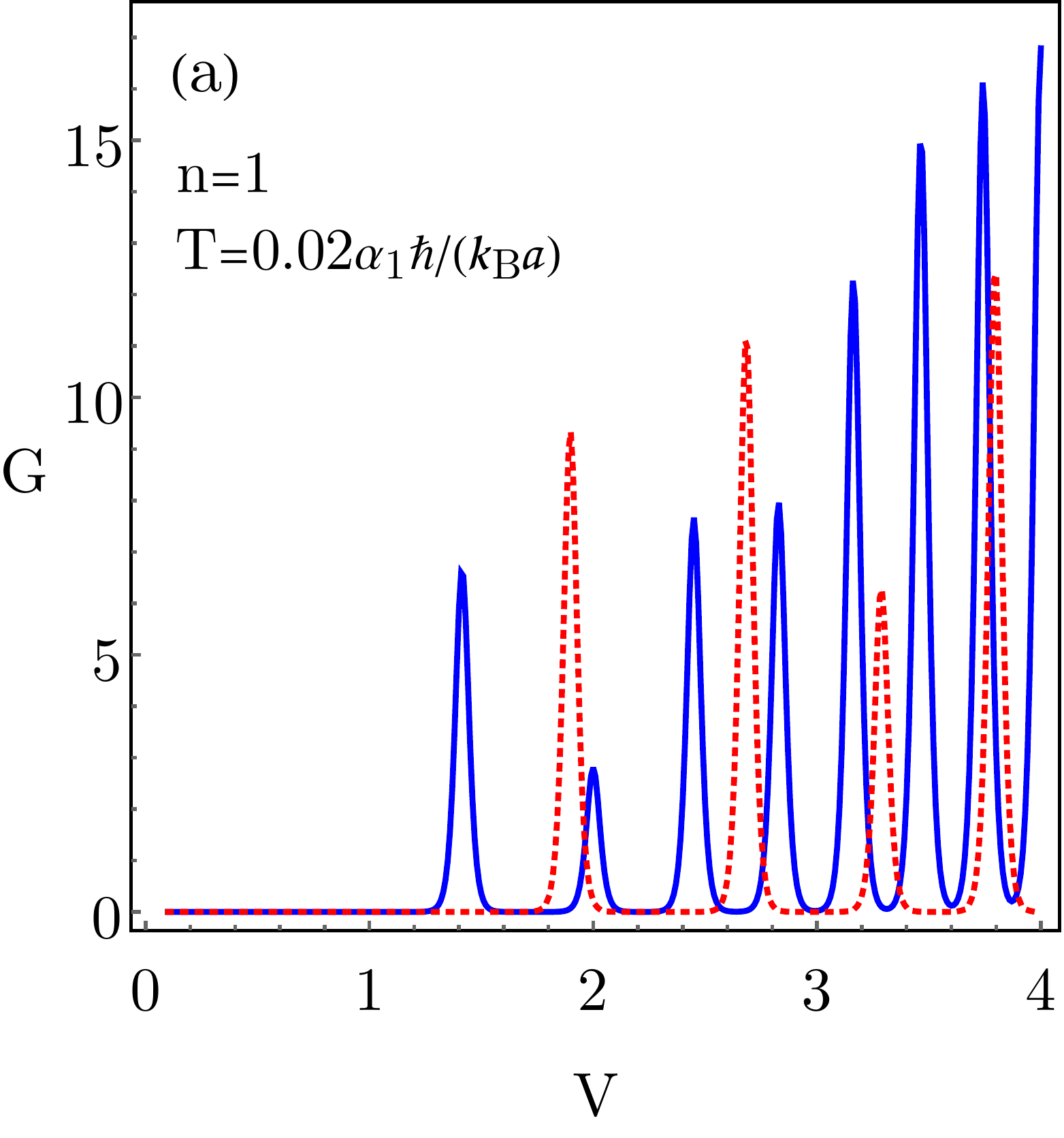}\label{fig2a} }
\hskip .1cm
{\includegraphics[width=0.318\textwidth]{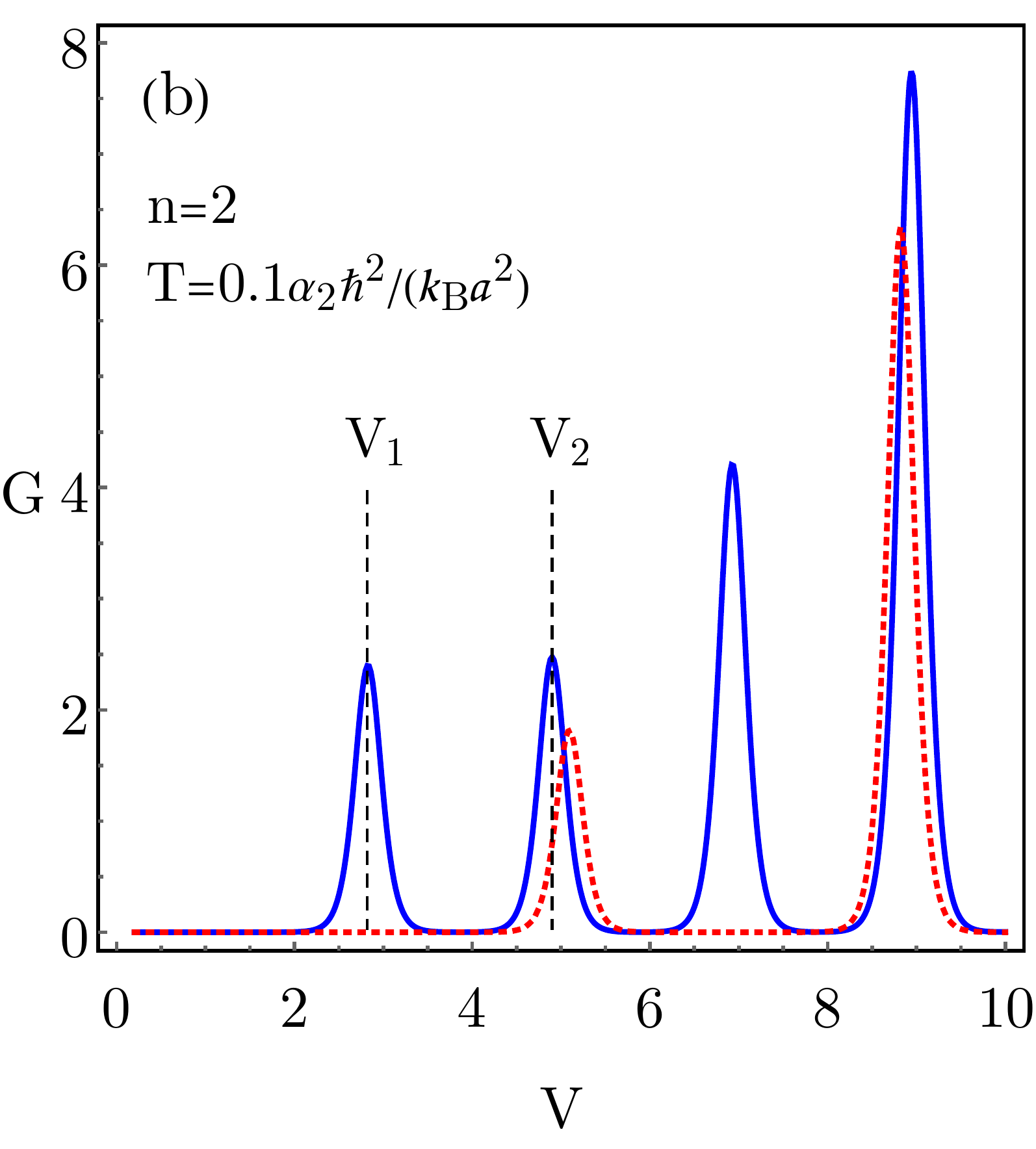}\label{fig2b}}
\hskip .1cm
{\includegraphics[width=0.325\textwidth]{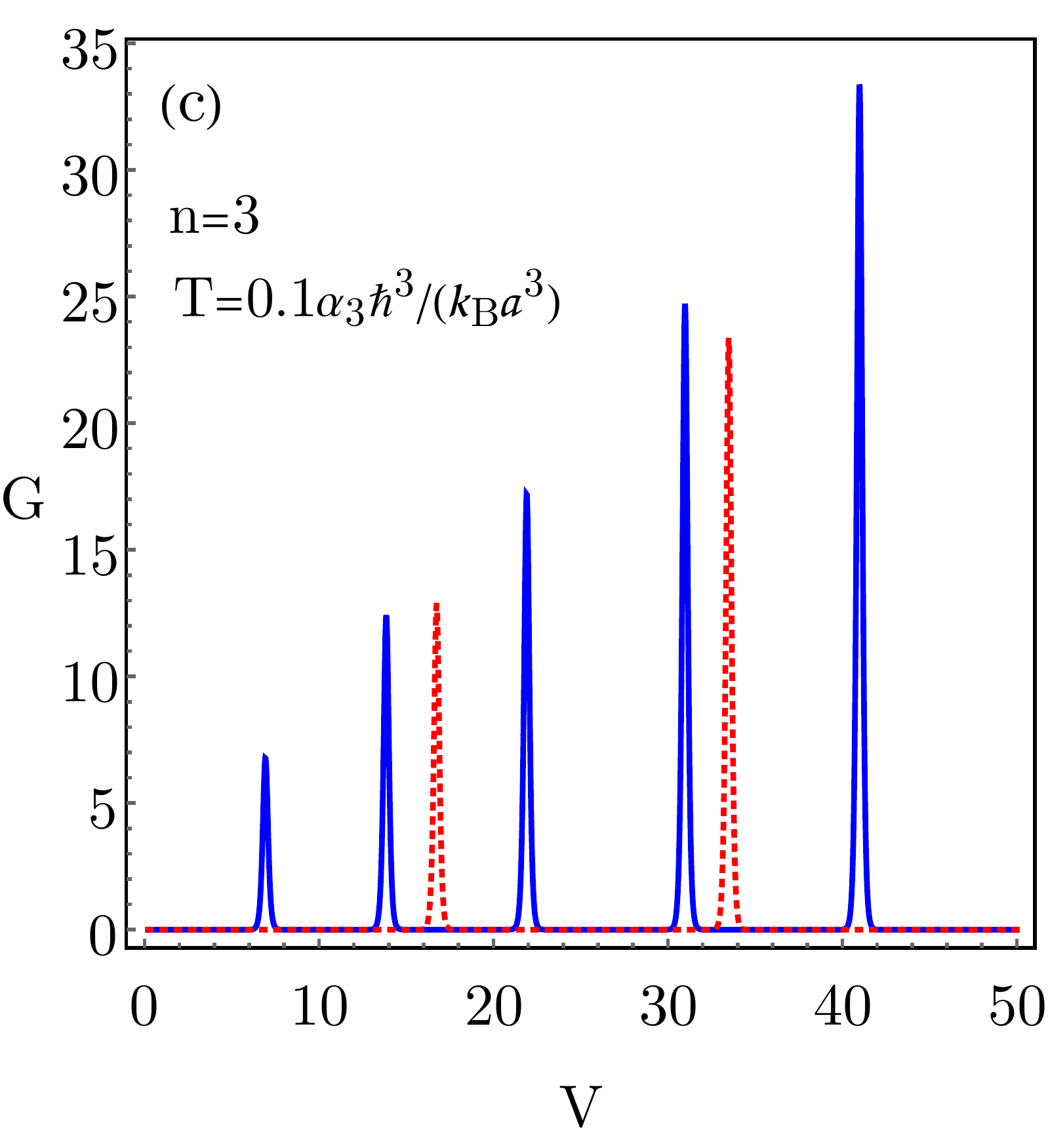}\label{fig2c}}
\caption{Conductance (dimensionless) calculated from the analytical Eq. (\ref{eq:conductance-final}), as a function of applied bias $V$ (in units of $\alpha_n(\hbar/a)^n$) at a finite temperature $T$. The solid blue (dotted red) line corresponds to a pseudo-magnetic flux $N_\phi=1$ ($N_\phi=1.8$). The panels (a), (b) and (c) correspond to the values of the monopole charge $n=1$, $n=2$ and $n=3$, respectively.}
\label{fig:conductances}
\end{figure*}

We here provide an affirmative answer to the above question by showing that the monopole charge $n$ is imprinted in the  scattering of the low-energy quasiparticles off a single dislocation defect, which may be measured in the electrical conductance, using the setup in Fig.~\ref{fig:setup}. Starting with  an effective low-energy description of the dislocation in terms of a flux tube with an effective $U(1)$ vector potential that minimally  couples to the low-energy nodal quasiparticles, we show that the electrical conductance displays  characteristic features directly related to the monopole charge. In particular, the ratio between the neighboring conductance peaks as a function of applied gate voltage can be used to directly infer this topological invariant, see Fig.~\ref{fig:conductances}. Furthermore, once the monopole charge is determined, the position of the conductance peak can be used to find the effective flux carried by the dislocation defect.

{\it Model.~}We start by considering a minimal Hamiltonian of a mWSM with broken time-reversal symmetry, describing the low-energy quasiparticles in the vicinity of two nodal points at the momenta $\xi {\bf K}_W=\xi(0,0,K_{W,z})$, with $\xi=\pm$ \cite{roy2017},
\begin{align}\label{eq:MWSM_Hamiltonian}
H^{(n)}_\xi({\bf k})&=\alpha_nk^n_\bot\left[\sigma_x\cos(n\phi_k)+\sigma_y\sin(n\phi_k) \right]+ \xi v_zk_z\sigma_z
\end{align}
where  $k_\bot^2=k_x^2+k_y^2$, $\phi_k=\arctan\left(k_y/k_x \right)$, and $n>0$. $\alpha_n$ and $v_z$ are the parameters of the model. Notice that while $\alpha_1$ has dimensions of velocity, $\alpha_2$ has dimensions of inverse mass. The anticommuting property of Pauli matrices implies the dispersion of the low-energy quasiparticles in the form
\begin{equation}
    \epsilon_{n,\vec{k}}^{\lambda}=\lambda\sqrt{\alpha^2_nk^{2n}_\bot+v_z^2k_z^2}\equiv \lambda\epsilon_{n,\mathbf{k}},
    \label{eq:eigenstate0}
\end{equation}
where $\lambda=\pm$ labels conduction and valence band, respectively.
The corresponding eigenstates, satisfying $H_{\xi}^{(n)}\Psi^{(\lambda,\xi)}(\mathbf{r}) =  \lambda\epsilon_{n,\mathbf{k}} \Psi^{(\lambda,\xi)}(\mathbf{r})$,  read
\begin{eqnarray}
\Psi^{(\lambda,\xi)}(\mathbf{r}) = \frac{e^{i\mathbf{k}\cdot{\mathbf{r}}}}{\sqrt{2}}\displaystyle\left(\begin{array}{cc}
\chi_+\\
\lambda \chi_- e^{i n \phi}
\end{array}
\right),
\label{eq:free_spinor}
\end{eqnarray}
where $\chi_\mu\equiv \chi_\mu(\lambda,\xi,v_z,k_z,\epsilon_{n,\vec{k}})=\sqrt{1+\mu\lambda\xi(v_z k_z/\epsilon_{n,\vec{k}})}$, and $\mu=\pm$.

{\it Multi-Weyl semimetal in the presence of a screw dislocation defect.~} We now consider a single screw dislocation defect with the Burgers vector oriented in the $z$ direction. Such a dislocation is described as a tube carrying an effective pseudo-magnetic (time-reversal preserving) flux, $\Phi_D=\Phi_D^{\rm top}+\Phi_D^{\rm mat}$, with both topological $\Phi_D^{\rm top}\sim {\bf K}_W\cdot{\bf b}$ ~\cite{juricic2012} and non-topological (material specific) part $\Phi_D^{\rm mat}\sim \beta |{\bf b}|$, where ${\bf b}$ is the Burgers vector of the dislocation and $\beta$ is an effective Gruneisen parameter of the material~\cite{Cortijo_015,chernodub2017}. To account for the physics at the scale of the defect core within the continuum theory, we employ a regularization so that the effective magnetic field ${\bf B}_D = B_D {\bf e}_z$ is concentrated within the radius $a$. Furthermore, the flux tube extends in the $z$-direction along the length $L$ and the effective field is assumed to be constant, $B_D=\Phi_D/\pi a^2$, see also Fig.~\ref{fig:setup}.
The low energy fermionic excitations in the vicinity of the two nodal points are minimally coupled to the  emergent chiral vector potential as ${\bf k}\rightarrow {\bf k}+\xi{\bf A}_D$ (hereafter, we set $e=\hbar=k_B=1$). Long-range effect of this gauge field is then encoded through the matching condition between the scattered waves and the defect induced Landau levels (LLs). This procedure is justified  {\it a posteriori} since the radius $a$ enters only as a prefactor in the transmission function, while the remaining expression is a function of only the monopole charge and the dimensionless effective pseudo-magnetic flux of the defect, see Eq.~\eqref{eq:conductance-final} and the discussion there. In turn, the features in the conductance allowing the direct probing of the monopole charge do not depend on the defect core radius.

We start by solving the Landau problem for the continuum Hamiltonian, given by Eq.~(\ref{eq:MWSM_Hamiltonian}),
\begin{equation}\label{eq:mWSM-Hamiltonian-B}
H^{(n)}_{\xi}({\bf k}+\xi{\bf A}_D)|\Psi_{m}^{(\lambda,\xi)}\rangle=\lambda E_{m}|\Psi^{(\lambda,\xi)}_m\rangle,
\end{equation}
with ${\bf A}_D=(B_D/2)(-y{\bf e}_x+x{\bf e}_y)$ corresponding to a constant effective pseudo-magnetic field $B_D$  carried by the defect.
Energy of the dislocation induced  LLs  then reads
\begin{equation}\label{eq:PLL}
E_{m}=\sqrt{(v_zk_z)^2+\left(\frac{2\alpha_n^{2/n}}{\ell^2}\right)^n\frac{m!}{(m-n)!}},
\end{equation}
where $m\geq0$ is an integer LL index and $\ell=1/\sqrt{B_D}$ is the magnetic length of the order of the dislocation core radius. Notice that at $k_z=0$ there are precisely $n$ zero-energy states, while the finite energy ones ($m\geq n$) scale as $E_m\sim\sqrt{m!/(m-n)!}$. We display the form of only the non-zero LLs (with finite energy at $k_z=0$) as these are the ones contributing to the scattering
\begin{eqnarray}
&&\Psi^{(\lambda,\xi=+)}_{m,M}({\bf r})=\frac{e^{-r_\ell^2/2}e^{ik_zz}}{\sqrt{4\pi L \ell^2 (|M|+ m -n/2)!}}\nonumber\\
&\times&\left(\begin{array}{c}\chi_+ C(m-n,r_\ell,|M|+n/2) e^{i(M-\frac{n}{2})\phi}\\
\lambda \chi_- C(m,r_\ell,|M|-n/2) e^{i( M+\frac{n}{2})\phi}
\end{array} \right),
\label{eq:LL1}
\end{eqnarray}
where the functions
$\chi_\mu\equiv \chi_\mu(\lambda,\xi=+,v_z,k_z,E_m)$, $C(m,r_\ell,n)=\sqrt{m!}(-i)^m r_\ell^n L_m^n(r_\ell^2)$, with $r_\ell\equiv r/\ell\sqrt{2}$, $L_m^n(x)$ as the associated Laguerre polynomial, and the node index $\xi=+$, while $\Psi^{(\lambda,\xi=-)}_{m,M}=\sigma_x \Psi^{(\lambda,\xi=+)}_{m,M}$. Here, $M=m'+n/2$, with $m'$ an integer, is the eigenvalue of the total angular momentum operator $J_z=L_z+(n/2)\sigma_z$ for the multi-Weyl fermion with monopole charge $n$, $[J_z,H^{(n)}_\xi({\bf k}+\xi{\bf A}_D)]=0$, where the Hamiltonian is given by Eq.~(\ref{eq:MWSM_Hamiltonian}), see the Supplementary Materials (SM), Sec.~S1~\cite{SM}.


{\it Scattering analysis.~} We now compute the scattering amplitude for the incident waves propagating along the $x$-axis and scattering off the dislocation defect within an approach analogous as in Refs.~\cite{Munoz-2017,Soto_018}. To find the imprint of the effective $U(1)$ gauge field emerging from the dislocation defect on the electronic states scattering off it, as a first step, we impose the continuity condition at the dislocation core
\begin{equation}\label{eq:matching}
\Psi_{out,M}^{(\lambda,\xi)}(r = a,\phi,z) = \Psi_{m,M}^{(\lambda,\xi)}(r=a,\phi,z),
\end{equation}
with  $\Psi_{m,M}^{(\lambda,\xi)}(r,\phi,z)$ as the LL state given by Eq.~(\ref{eq:LL1}), while the outgoing wave is parametrized in terms of the scattering phase shifts as

\begin{equation}
\Psi_{out,M}^{(\lambda,\xi)}({\bf r}) = {\tilde C}_{M}e^{i k_z z}
\left(\begin{array}{c} \varphi_{M-n/2}(\delta_{M},{r})\\
\frac{i^n\alpha_n k_{\perp}^n}{\xi v_z k_z + \lambda \epsilon_{n,\mathbf{k}}}\,\,
\varphi_{M+n/2}(\delta_{M},{r})
\end{array}\right),
\label{eq:out-1}
\end{equation}

with

\begin{align}
\varphi_{M\pm n/2}(\delta_{M},r)&= e^{i(M \pm n/2)\phi}
[\cos\delta_{M} J_{M\pm n/2}(k_{\perp}r)\nonumber\\
&-\sin \delta_{M} Y_{M \pm n/2}(k_{\perp}r)],
\end{align}
 where $J_m(x)$ and $Y_m(x)$ are the Bessel functions of the first and the second kind, respectively.
The matching condition in Eq.~(\ref{eq:matching}) leads to the form of the phase shifts $\delta_M$, whose exact analytical expressions are provided in the SM~\cite{SM}, Eqs.~(S59) and (S60).
Finally, the asymptotic form of the scattered waves far away from the defect ($r\rightarrow\infty$)
\begin{eqnarray}
\left.\Psi^{(\lambda,\xi)}_{out}\right|_{r\rightarrow\infty}= \left.\left[\Psi^{(\lambda,\xi)}_{inc} + \left(\begin{array}{c}f_1(\phi)\\
f_2(\phi)
\end{array}\right)\frac{e^{i k_{\perp}r + i k_z z}}{\sqrt{r}}\right]\right|_{r\rightarrow\infty}
\label{eq:out}
\end{eqnarray}
determines both the constants ${\tilde C}_{M}=\exp(i\delta_{M}+i m\pi/2)/\sqrt{2}$ in the scattered wave in Eq. (\ref{eq:out-1}) and the  scattering amplitudes $f_{1,2}(\phi)$, the form of which are explicitly given in Sec.~S2 of the SM~\cite{SM}. Notice that the phase shifts explicitly  depend on the monopole charge, $n$. Next, we compute the conductance to show the signatures of the monopole charge in this transport observable.


{\it Transmission and Landauer conductance.~} Following a Landauer ballistic approach~\cite{Munoz-2017,Soto_018}, the average  transmission function in the $x$ direction for the states scattered off the dislocation (see Fig.~\ref{fig:setup}) is obtained as
\begin{eqnarray}
{\bar T}_{\xi}(E_m) = \int_{-\pi/2}^{\pi/2}d\phi\cos\phi\left(\frac{1}{\sigma(E_m)}\right)\left.\frac{d\sigma(\phi)}{d\phi}\right|_{k_\perp^n=\frac{E_m}{\alpha_n}}.\nonumber\\
\label{eq:transmission}
\end{eqnarray}
Here, the elastic scattering is explicitly enforced by the condition $k_\perp=(E_m/\alpha_n)^{1/n}$ on the right-hand side, 
$\sigma(E_m)=4(\alpha_n/E_m)^{1/n}\sum_{M\in\mathbb{Z}}|F(\delta_M)|^2$, with the LLs energy $E_m$ at $k_z=0$ given in Eq.~\eqref{eq:PLL}. The effective cross section in the angular in-plane direction reads
\begin{equation}
\frac{d\sigma(\phi)}{d\phi}=\frac{2}{\pi k_\perp}\sum_{M,M'\in \mathbb{Z}} F(\delta_M,\phi)F^*(\delta_{M'},\phi),
\end{equation}
where $F(\delta_M,\phi)=\exp(i\delta_M+i M \phi)\sin\delta_M$, see Sec.~S3 of the SM for details~\cite{SM}.

The total current
through the junction is given by the sum of the individual ones from each of the Weyl nodes $\xi = \pm$,
$I = I_{+} + I_{-}$. Finally, assuming that both contacts are held at the same temperature $T_L = T_R = T$, but at different chemical potentials $\mu_L = \mu_R + V$, $\mu_{L,R}>0$,  with the difference given by the gate voltage $V$, the electrical conductance 
$G(T,V)=\partial I/\partial V$ is given by the expression
\begin{equation}\label{eq:conductance-final}
G(T,V)= \frac{ \alpha_{n}^{1/n}}{T}\sum_{\lambda,\xi,m}\mathcal{T}(E_m) f_L(\lambda E_m)
\left[1 - f_L(\lambda E_m) \right],
\end{equation}
where we used that the transmission function in Eq.~\eqref{eq:transmission} is independent of the node, and $f_{L,R}(E)$ are the local Fermi-Dirac distribution functions at the contacts.
The transmission function $\mathcal{T}(E_m)$ is obtained by performing the integration over the angle in Eq.~\eqref{eq:transmission}, with the explicit form in terms of the phase shifts given by Eq.~(S74) in the SM~\cite{SM}. It is important to notice that the effective transmission function entering the conductance given in Eq.~\eqref{eq:conductance-final}, up to a prefactor is a function of only the effective flux $N_\phi$ and the monopole charge $n$, and independent of the radius of the dislocation core $a$; see also Sec.~S4 in the SM~\cite{SM}. This feature is fundamentally important for the probing of the monopole charge in the setup we propose, as discussed below.  


%


{\it Results.~} The differential conductance as a function of the applied voltage is displayed in Fig.~\ref{fig:conductances}, for the cases $n=1,2,3$, respectively, at finite temperature and the dislocation defect characterized by the dimensionless pseudo-magnetic flux $N_{\phi}$. A remarkable feature in Fig.~\ref{fig:conductances} is the presence of the peaks corresponding to resonances located at each dislocation-induced LL $E_m$ ($m\ge n$, see Eq.~(\ref{eq:PLL})), i.e. the voltage $V_{k} \sim E_{m = n + k}$, for $k = 1,2,\ldots$ consecutive peaks. Those resonances are expected, since they correspond to the states that can be transmitted through the dislocation region according to Eq.~(\ref{eq:conductance-final}). As can be seen in Fig.~\ref{fig:conductances}, because of the form of LLs in  Eq.~(\ref{eq:PLL}), the position of a resonance peak $\sim(N_{\phi})^{n/2}$, and hence can in principle be used to determine the magnitude of the dislocation pseudo-magnetic flux.

The monopole charge of the semimetal can be found directly from the scaling of the conductance as a function of the applied voltage for a fixed value of the flux $N_\phi$, shown in  Fig. \ref{fig:conductances}. Since the LLs at non-zero energy and for $k_z=0$ arise for $m \ge n$, as given in Eq. (\ref{eq:PLL}), the first two peaks in the conductance as a function of the bias voltage are resonant with the first two LLs  at non-zero energy, $V_1 \sim  E_{m=n}$ and $V_2 \sim  E_{m = n+1}$, respectively. Therefore, the value of $n$ is directly related to the ratio of the first two peaks in the conductance:
\begin{eqnarray}
n = \left\lfloor\left(\frac{V_{2}}{V_1}\right)^{2}-1\right\rceil.
\label{eq:voltage_peaks}
\end{eqnarray}
Here, the symbol $\lfloor x\rceil$ represents the nearest integer to $x$. Take for instance the calculated conductance vs bias voltage curve  displayed in Fig.~\ref{fig:conductances}(b), where we can directly read off the first and second peak, with the precision at two significant figures, to be at $V_1 \simeq 2.9$, and $V_2 \simeq 5.0$, respectively. Now applying Eq.~\eqref{eq:voltage_peaks}, we obtain (with the same precision) $n \simeq 2.0$, in an excellent agreement with the monopole charge considered in this example. Furthermore, the number of the effective magnetic flux quanta (in units of $hc/e=2\pi$) associated with the dislocation defect can be calculated directly from Eq.~\eqref{eq:PLL}.  Since $E_{m = n} = V_1$, we find
\begin{eqnarray}
N_{\phi} = \frac{1}{2}\left(\frac{V_1}{\sqrt{n!}}\right)^{2/n},
\label{eq:flux_voltage}
\end{eqnarray}
which in the same example yields $N_\phi\simeq 2$ again in agreement with the exact value chosen there. These features are directly related to the independence of the transmission function of the dislocation radius. Otherwise, the maxima in the conductance would acquire an additional dependence on the dislocation radius, which would spoil their universal features. 


{\it Discussion \& Outlook.~} To summarize, in this paper we propose that a dislocation, rather ubiquitous lattice defect in crystals, can serve as a direct probe of the monopole charge of a generalized Weyl semimetal in a simple mesoscopic setup, illustrated in Fig.~\ref{fig:setup}. A dislocation defect acting as an emergent $U(1)$ gauge field leaves its imprint on the spectrum and the scattering phase shifts, which in turn yield the characteristic conductance peaks, as shown in Fig.~\ref{fig:conductances}. The ratio of the neighboring peaks, as we showed, can be used to directly determine the monopole charge, while the position of the peak in the $G-V$ plane directly yields the value of the effective $U(1)$ flux carried by the defect, see Fig.~\ref{fig:conductances}(b). 

We remark that in principle, a real magnetic field would induce similar qualitative spectral and transport properties as the dislocation considered here, since Landau levels would also emerge under these conditions. However, in practice it is nearly impossible to impose an external magnetic field confined to a localized flux tube as would be required here, unless perhaps the sample was placed right on top of a vortex in a type II superconducting material. Another conceptually interesting configuration is given by the combination of a confined external magnetic field and a dislocation, which may yield the breaking of the symmetry between the Weyl nodes, thus leading to chirality-dependent  transport.

Our results therefore provide a direct guidance for the experimental detection of the monopole charge in the candidate double-Weyl ($n=2$) materials, e.g. HgCr$_2$Sr$_4$~\cite{XuPRL2011,FangPRL2012} and SrSi$_2$~\cite{Huang2016,Singh2018}, and in the triple-Weyl ($n=3$) compounds, for instance, the proposed transition-metal monochalcogenides A(MoX$_3$) with A=Na, K, Rb, In, Tl, and X=S, Se, Te~\cite{Zunger2017}, and BaAgAs \cite{Mardanya2019}; see also Refs. \cite{Bernevig2018,Gao2019}. The proposed setup may serve as a complement to the existing probes, such as angle resolved photoemission spectroscopy (ARPES).

It is important to point out that the transmission function in our setup, where the electronic states carry $k_z\approx0$, depends only on the effective flux and the monopole charge. When the effects of the propagating modes along the $z$-direction are included, the location of the conductance peaks will display an extra dependence on the velocity $v_z$, according to Eq.~\eqref{eq:PLL}, with $1/L$ as the momentum scale along the $z$-axis. Therefore, in this case two pairs of nearest neighbor peaks are enough to determine the value of the monopole charge. On the other hand, one may then estimate the effect of a finite sample size in the $z$-direction on the extracted value of the monopole charge, $n$. Finally, we emphasize that the $x-y$ plane  does not feature Fermi arc surface states, and therefore they do not contribute to the conductance calculated here. However, it remains as an interesting problem to consider on general grounds the interplay of the dislocation and the Fermi arc contributions in the transport.

Our results motivate further studies of the role of topological defects in the quantum  transport in topological nodal materials. In particular, the effects of many dislocation defects, whcih are expected to only shift position and the height of the conductance peaks, as well as of disclinations  and  grain  boundaries,  should  be  considered. Finally, we anticipate that our results will motivate experimental and numerical studies of the effects discussed in this work.


{\it Acknowledgements.~}V. J. is grateful to Cosma Fulga for fruitful discussions. This work was supported by Fondecyt Grants 1190361 and 1200399, and by ANID PIA Grant No. ACT192023. V.J. acknowledges the support of the Swedish Research Council (Grant No. VR 2019-04735).

\bibliography{references}

\end{document}